\documentclass[dvips]{article}
\usepackage{supertabular,lscape,epsfig}
\usepackage{amssymb}
\usepackage{amsmath}
\usepackage{rotating}
\DeclareSymbolFont{ppa}{OT1}{ppl}{m}{it}
\DeclareMathSymbol{\vv}{\mathalpha}{ppa}{'166}

\begin{document}

\newcommand{\dd}{\,{\rm d}}
\newcommand{\ie}{{\it i.e.},\,}
\newcommand{\etal}{{\it et al.\ }}
\newcommand{\eg}{{\it e.g.},\,}
\newcommand{\cf}{{\it cf.\ }}
\newcommand{\vs}{{\it vs.\ }}
\newcommand{\zdot}{\makebox[0pt][l]{.}}
\newcommand{\up}[1]{\ifmmode^{\rm #1}\else$^{\rm #1}$\fi}
\newcommand{\dn}[1]{\ifmmode_{\rm #1}\else$_{\rm #1}$\fi}
\newcommand{\upd}{\up{d}}
\newcommand{\uph}{\up{h}}
\newcommand{\upm}{\up{m}}
\newcommand{\ups}{\up{s}}
\newcommand{\arcd}{\ifmmode^{\circ}\else$^{\circ}$\fi}
\newcommand{\arcm}{\ifmmode{'}\else$'$\fi}
\newcommand{\arcs}{\ifmmode{''}\else$''$\fi}
\newcommand{\MS}{{\rm M}\ifmmode_{\odot}\else$_{\odot}$\fi}
\newcommand{\RS}{{\rm R}\ifmmode_{\odot}\else$_{\odot}$\fi}
\newcommand{\LS}{{\rm L}\ifmmode_{\odot}\else$_{\odot}$\fi}

\newcommand{\Abstract}[2]{{\footnotesize\begin{center}ABSTRACT\end{center}
\vspace{1mm}\par#1\par
\noindent
{~}{\it #2}}}

\newcommand{\TabCap}[2]{\begin{center}\parbox[t]{#1}{\begin{center}
  \small {\spaceskip 2pt plus 1pt minus 1pt T a b l e}
  \refstepcounter{table}\thetable \\[2mm]
  \footnotesize #2 \end{center}}\end{center}}

\newcommand{\TableSep}[2]{\begin{table}[p]\vspace{#1}
\TabCap{#2}\end{table}}

\newcommand{\FigCap}[1]{\footnotesize\par\noindent Fig.\  %
  \refstepcounter{figure}\thefigure. #1\par}

\newcommand{\TableFont}{\footnotesize}
\newcommand{\TableFontIt}{\ttit}
\newcommand{\SetTableFont}[1]{\renewcommand{\TableFont}{#1}}

\newcommand{\MakeTable}[4]{\begin{table}[htb]\TabCap{#2}{#3}
  \begin{center} \TableFont \begin{tabular}{#1} #4
  \end{tabular}\end{center}\end{table}}

\newcommand{\MakeTableSep}[4]{\begin{table}[p]\TabCap{#2}{#3}
  \begin{center} \TableFont \begin{tabular}{#1} #4
  \end{tabular}\end{center}\end{table}}

\newenvironment{references}%
{
\footnotesize \frenchspacing
\renewcommand{\thesection}{}
\renewcommand{\in}{{\rm in }}
\renewcommand{\AA}{Astron.\ Astrophys.}
\newcommand{\AAS}{Astron.~Astrophys.~Suppl.~Ser.}
\newcommand{\ApJ}{Astrophys.\ J.}
\newcommand{\ApJS}{Astrophys.\ J.~Suppl.~Ser.}
\newcommand{\ApJL}{Astrophys.\ J.~Letters}
\newcommand{\AJ}{Astron.\ J.}
\newcommand{\IBVS}{IBVS}
\newcommand{\PASP}{P.A.S.P.}
\newcommand{\Acta}{Acta Astron.}
\newcommand{\MNRAS}{MNRAS}
\renewcommand{\and}{{\rm and }}
\section{{\rm REFERENCES}}
\sloppy \hyphenpenalty10000
\begin{list}{}{\leftmargin1cm\listparindent-1cm
\itemindent\listparindent\parsep0pt\itemsep0pt}}%
{\end{list}\vspace{2mm}}

\def\TYLDA{~}
\newlength{\DW}
\settowidth{\DW}{0}
\newcommand{\dw}{\hspace{\DW}}

\newcommand{\refitem}[5]{\item[]{#1} #2%
\def\REFARG{#3}\ifx\REFARG\TYLDA\else, {\it#3}\fi
\def\REFARG{#4}\ifx\REFARG\TYLDA\else, {\bf#4}\fi
\def\REFARG{#5}\ifx\REFARG\TYLDA\else, {#5}\fi.}

\newcommand{\Section}[1]{\section{#1}}
\newcommand{\Subsection}[1]{\subsection{#1}}
\newcommand{\Acknow}[1]{\par\vspace{5mm}{\bf Acknowledgments.} #1}
\pagestyle{myheadings}

\newfont{\bb}{ptmbi8t at 12pt}
\newcommand{\xrule}{\rule{0pt}{2.5ex}}
\newcommand{\xxrule}{\rule[-1.8ex]{0pt}{4.5ex}}
\def\thefootnote{\fnsymbol{footnote}}

\begin{center}
{\Large\bf Eclipsing Binary Stars in the OGLE-III\\
Galactic Disk Fields\footnote{Based on observations obtained with the
1.3-m Warsaw telescope at the Las Campanas Observatory
of the Carnegie Institution for Science.}}
\vskip1cm
{\bf
P.~~P~i~e~t~r~u~k~o~w~i~c~z$^1$,~~P.~~M~r~\'o~z$^1$,~~I.~~S~o~s~z~y~\'n~s~k~i$^1$,\\
~~A.~~U~d~a~l~s~k~i$^1$,~~R.~~P~o~l~e~s~k~i$^{1,2}$,~~M.~K.~~S~z~y~m~a~\'n~s~k~i$^1$,\\
~~M.~~K~u~b~i~a~k$^1$,~~G.~~P~i~e~t~r~z~y~\'n~s~k~i$^{1,3}$,~~{\L}.~~W~y~r~z~y~k~o~w~s~k~i$^{1,4}$,\\
~~K.~~U~l~a~c~z~y~k$^1$,~~S.~~K~o~z~{\l}~o~w~s~k~i$^1$ and~~J.~~S~k~o~w~r~o~n$^1$\\}
\vskip3mm
{
$^1$Warsaw University Observatory, Al. Ujazdowskie 4, 00-478 Warszawa, Poland\\
e-mail:
(pietruk,pmroz,soszynsk,udalski,rpoleski,msz,mk,pietrzyn,\\
wyrzykow,kulaczyk,simkoz,jskowron)@astrouw.edu.pl\\
$^2$ Department of Astronomy, Ohio State University, 140 W. 18th Ave.,\\
Columbus, OH 43210, USA\\
$^3$ Universidad de Concepci{\'o}n, Departamento de Fisica,\\
Casilla 160-C, Concepci{\'o}n, Chile\\
$^4$ Institute of Astronomy, University of Cambridge, Madingley Road,\\
Cambridge CB3 0HA, UK\\
}
\end{center}

\Abstract{We present the analysis of 11~589 eclipsing binary stars identified
in twenty-one OGLE-III Galactic disk fields toward constellations of Carina,
Centaurus, and Musca. All eclipsing binaries but 402 objects are new
discoveries. The binaries have out-of-eclipse brightness between $I=12.5$~mag
and $I=21$~mag. The completeness of the catalog is estimated at a level
of about 75\%. Comparison of the orbital period distribution for the
OGLE-III disk binaries with systems detected in other recent large-scale
Galactic surveys shows the maximum around 0.40~d and an almost flat
distribution between 0.5~d and 2.5~d, indepedent of population. Ten
doubly eclipsing systems and one eclipsing-ellipsoidal object were found
among thousands of variables. Nine of them are candidates for
quadruple systems. We also identify ten eclipsing subdwarf-B type
binary stars and numerous eclipsing RS CVn type variables.
All objects reported in this paper are part of the OGLE-III Catalog
of Variable Stars.}

{Galaxy: disk -- binaries: eclipsing -- variables: general -- starspots}


\Section{Introduction}

Binaries are common among stars of our universe. It is believed that two out
of every three stars in our Galaxy are in a binary or a multiple system
and a few per cent of main sequence binaries could be observed as
eclipsing systems (S\"oderhjelm and Dischler 2005). Eclipsing binaries
are a powerful tool in astrophysical diagnostics. They are widely used as
probes of stellar structure and evolution. Eclipsing systems allow
a direct determination of the fundamental parameters (such as masses and
radii) of the component stars to high accuracy (\eg Pietrzy\'nski
\etal 2010, 2012) and provide excellent tests of stellar
models (\eg Ribas \etal 2000, Morales \etal 2010). Binaries serve as
superb distance indicators to nearby galaxies (\eg Bonanos \etal 2006,
North \etal 2010, Vilardell \etal 2010, Pietrzy\'nski \etal 2013)
and help to trace the structure and evolution of the Milky Way
(\eg Nataf \etal 2012, He{\l}miniak \etal 2013).
Precize measurements of orbital motion of a binary can be used to
search for orbiting companions through the light-travel time effect
and transits (\eg Doyle \etal 2011, Welsh \etal 2012).

Following the advent of wide-field photometric surveys, the number of new
eclipsing variables has increased rapidly. The Optical Gravitational
Lensing Experiment (OGLE) is a long-term project with the main aim to
detect microlensing events toward the Galactic bulge (Udalski \etal 1993,
Udalski 2003). However, regular observations of the Milky Way
and Magellanic Cloud stars, conducted in some fields
for over 21 years, allow us to discover and explore the variety
of variable objects. The OGLE collection of classified variable stars
recently has exceeded 400 000 objects. Most of them are
pulsating red giants such as Long Period Variables ($\sim340 000$
objects, \eg Soszy\'nski \etal 2013), Classical Cepheids ($\sim8000$
objects, \eg Soszy\'nski \etal 2008), and RR Lyrae stars ($\sim 45 000$,
objects, \eg Soszy\'nski \etal 2011). Among binary stars, a large sample
of eclipsing systems ($\sim26 000$) was found in the Large Magellanic
Cloud (Graczyk \etal 2011).

This paper presents a catalog of 11 589 eclipsing stars detected
in twenty-one fields located in the Galactic disk toward constellations
of Carina, Centaurus, and Musca. The presented sample is a part of
the OGLE-III Catalog of Variable Stars. In the following sections
of this paper, we describe: the observations and reductions (Section 2),
the completeness of the search (Section 3), the catalog itself (Section 4),
likely X-ray counterparts to the optical detections (Section 5),
the brightness, amplitude, and period distributions (Section 6).
In Section 7, we focus on selected interesting objects and finally,
in Section 8, we summarize our results.


\Section{Observations and Data Reductions}

All the data presented in this paper were collected with the 1.3-m
Warsaw telescope at Las Campanas Observatory, Chile. The observatory
is operated by the Carnegie Institution for Science. During the
OGLE-III project, conducted in years 2001--2009, the telescope was
equipped with an eight-chip CCD mosaic camera with a scale of 0.26
arcsec/pixel and a field of view of $35\arcm \times 35\arcm$.
Details of the instrumentation setup can be found in Udalski (2003).

Twenty-one fields covering the total area of 7.12~deg$^2$ around the
Galactic plane between longitudes $+288\arcd$ and $+308\arcd$
were observed. Their location in the sky is shown in Fig.~1. The time
coverage as well as the number of data points obtained by the OGLE
project varies considerably from field to field (Fig.~2 and Table~1).
The vast majority of the observations, typically of 1500 to 2700 points per
field, were collected through the $I$-band filter with exposure times of 120~s
and 180~s. Additional observations, consisting of only 3--8 measurements,
were carried out in the $V$-band filter with an exposure time of 240~s.
Both filters closely resemble those of the standard Johnson-Cousins system.

\begin{table}
\centering
\caption{\small Basic data on monitored Galactic disk fields in OGLE-III}
\medskip
{\small
\begin{tabular}{lcccc}
\hline
Field  &  $l$ & $b$ & $N_{\rm nights}$ & $N_{{\rm exp},I}$~~ \\
\hline
CAR100~~  & $290\zdot\arcd6544$ & $-0\zdot\arcd7510$ & 202 & 2698 \\
CAR104~~  & $289\zdot\arcd8439$ & $-1\zdot\arcd7249$ & 161 & 1641 \\
CAR105~~  & $289\zdot\arcd2911$ & $-1\zdot\arcd9906$ & 167 & 1770 \\
CAR106~~  & $290\zdot\arcd5054$ & $-1\zdot\arcd6063$ & ~47 &~~896 \\
CAR107~~  & $288\zdot\arcd9089$ & $-2\zdot\arcd5647$ & 135 & 1530 \\
CAR108~~  & $288\zdot\arcd6343$ & $-2\zdot\arcd0343$ & 135 & 1513 \\
CAR109$^*$& $288\zdot\arcd4607$ & $-2\zdot\arcd9904$ & 145 & 1811 \\
CAR110~~  & $288\zdot\arcd1846$ & $-2\zdot\arcd4606$ & 135 & 1475 \\
CAR111~~  & $288\zdot\arcd3599$ & $-1\zdot\arcd5037$ & 134 & 1697 \\
CAR112~~  & $289\zdot\arcd0276$ & $-1\zdot\arcd4562$ & 134 & 1688 \\
CAR113~~  & $289\zdot\arcd5717$ & $-1\zdot\arcd1922$ & 134 & 1676 \\
CAR114~~  & $289\zdot\arcd3178$ & $-0\zdot\arcd6516$ & 134 & 1672 \\
CAR115$^*$& $288\zdot\arcd2783$ & $-3\zdot\arcd0629$ & 234 & 2001 \\
CAR116~~  & $288\zdot\arcd2176$ & $-3\zdot\arcd7841$ & 233 & 1963 \\
CAR117~~  & $288\zdot\arcd7274$ & $-3\zdot\arcd5017$ & 232 & 1922 \\
CAR118~~  & $288\zdot\arcd6602$ & $-4\zdot\arcd2207$ & 230 & 1888 \\
CEN106~~  & $293\zdot\arcd4598$ & $+0\zdot\arcd5784$ & ~44 &~~819 \\
CEN107~~  & $296\zdot\arcd2458$ & $+0\zdot\arcd1238$ & ~44 &~~815 \\
CEN108~~  & $307\zdot\arcd4281$ & $-1\zdot\arcd7417$ & 278 & 2374 \\
MUS100~~  & $305\zdot\arcd4335$ & $-2\zdot\arcd0928$ & 279 & 2468 \\
MUS101~~  & $306\zdot\arcd4749$ & $-2\zdot\arcd3261$ & 280 & 2428 \\
\hline
\noalign{\vskip3pt}
\multicolumn{5}{p{7.5cm}}{\footnotesize $^*$ -- Fields CAR109 and CAR115
significantly overlap with each other in the sky, but have no common
nights.}
\label{tab:obslog}
\end{tabular}}
\end{table}

\begin{figure}[htb]
\centerline{\includegraphics[angle=0,width=130mm]{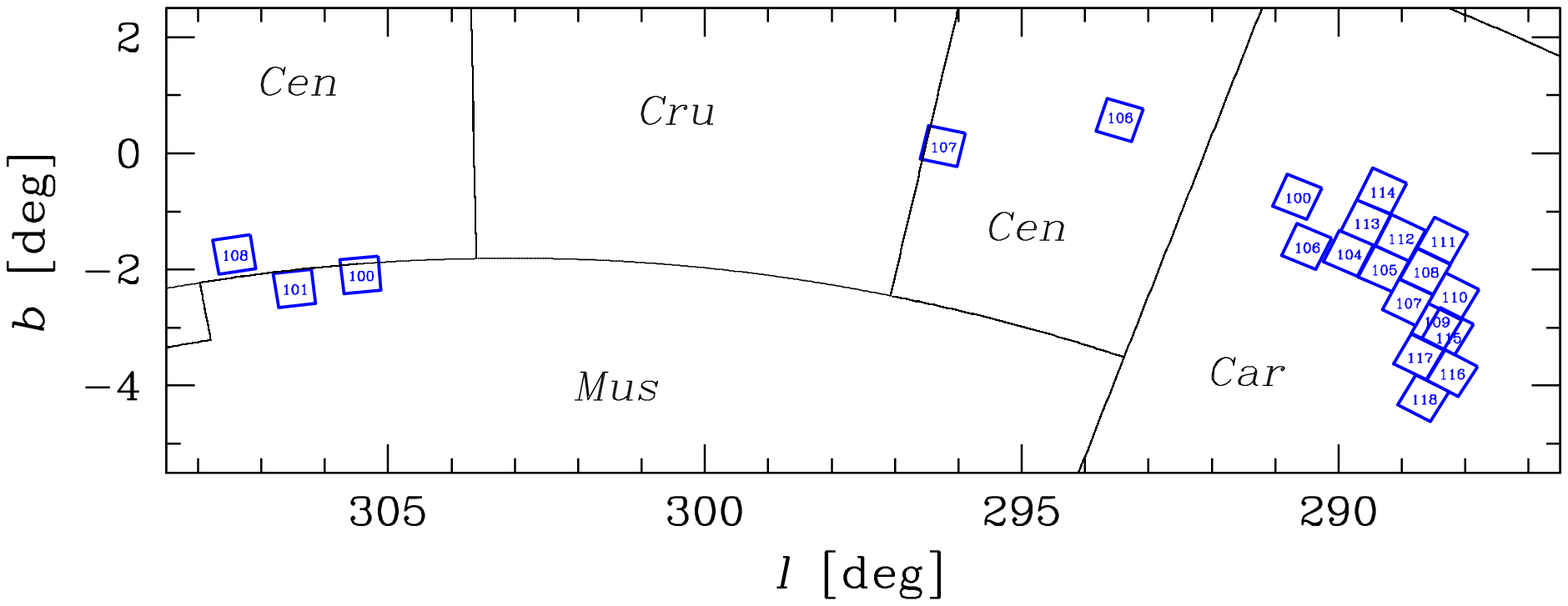}}
\FigCap{Location of the twenty-one OGLE-III disk fields in the Galactic
coordinates. Each field covers $35\arcm \times 35\arcm$ in the sky.
Two fields, CAR109 and CAR115, overlap with each other
in $\approx 67$\%. The total monitored area is 7.12~deg$^2$.}
\end{figure}

\begin{figure}[htb]
\centerline{\includegraphics[width=140mm]{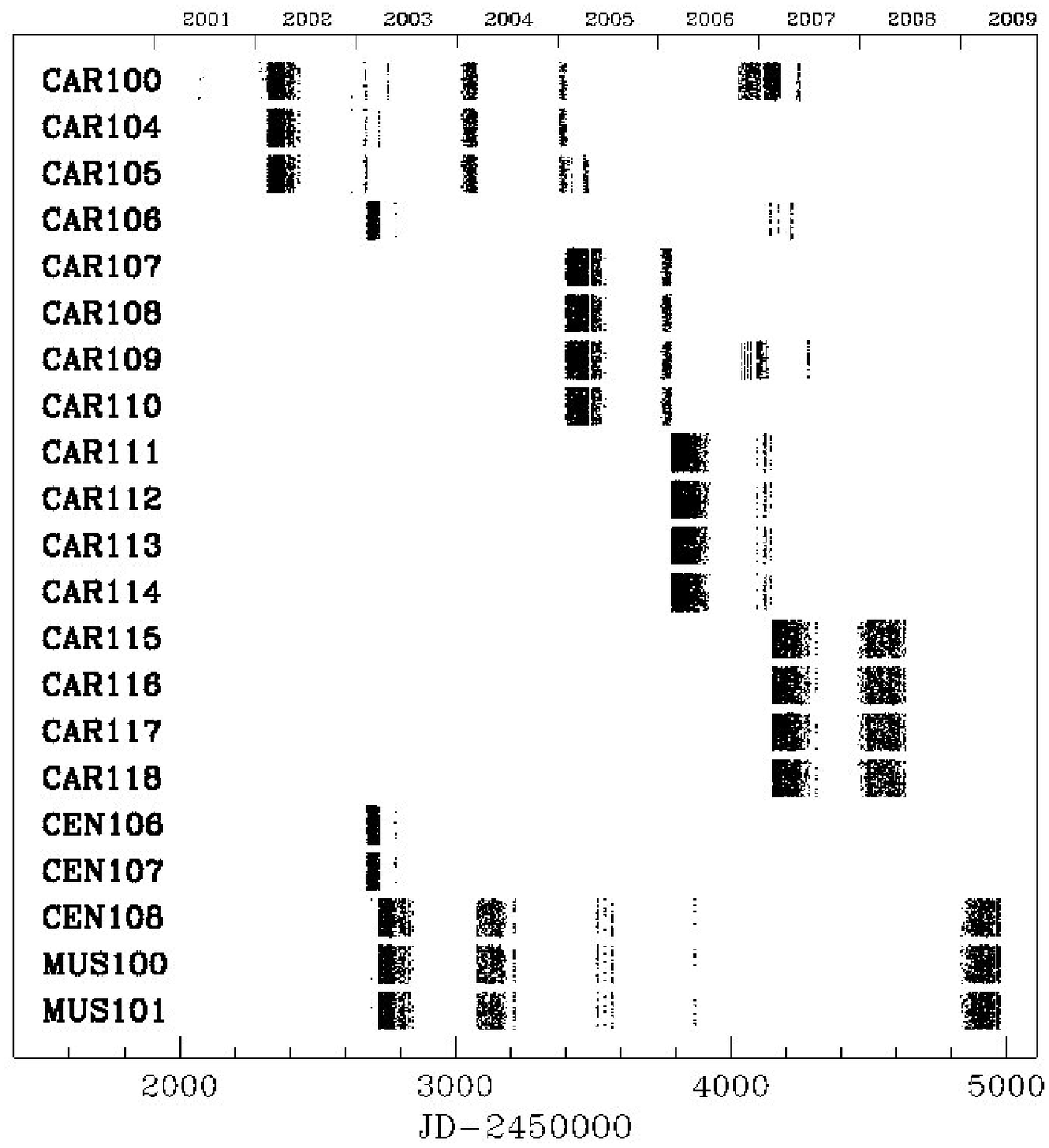}}
\FigCap{Graphical log of observations of the OGLE-III disk fields.
Each single point represents one exposure. To show irregular coverage
of the observations for each field we randomly spread the points
in the vertical direction. The number of points is between 815
and 2698 per field.}
\end{figure}

The time-series photometry was obtained with the standard OGLE data reduction
pipeline (Udalski \etal 2008, based on the Difference Image Analysis: DIA,
Alard and Lupton 1998, Wo\'zniak 2000). A period search was performed using the
F{\footnotesize NPEAKS} code (written by Z. Ko{\l}aczkowski, private communication)
for about $8.8 \times 10^6$ stars containing at least 30 points in the $I$-band
light curve. About 345~500 detections with a signal to noise S/N$>10$ were
visually checked for any kind of variability. The search resulted in about 13~000
eclipsing binary candidates, $\approx1000$ ellipsoidal-like candidates,
$\approx1000$ pulsating star candidates, and $\approx15~000$ miscellaneous
variables (mostly stars with spots). Details on the detected dwarf novae are
presented in Mr\'oz et al. (2013), while papers on other variables
(in particular on the pulsating stars) are in preparation.

In the next step, we rejected from the list of eclipsing binary candidates
false detections and artifacts that mimicked variability of neighboring real
eclipsing stars. Then, we manually removed outlying points from the light
curves of all eclipsing stars and improved orbital periods of binaries with
$P>1.0$~d. The uncertainty of the period is estimated at a level of about
10$^{-5}P$ for $P>1.0$~d. For stars with $P<1.0$~d their periods together
with errors were improved using the TATRY code (Schwarzenberg-Czerny 1996).
After the period correction we re-examined the list for the remaining false
period detections and for common variables independently detected
in overlapping fields. Combined light curves containing more data points
allowed us to additionally improve the orbital periods of the common stars.
We note that we were not able to estimate the orbital period
for 27 binaries with very small number of eclipses.

In the last reduction step, each variable was calibrated from the
instrumental to the standard magnitudes. For stars with available
measurements in the $V$ band we evaluated $V-I$ colors around
corresponding phases in the $I$-band light curve. Then the mean color
was calculated after rejection of two most outlying values.
In the case of red stars ($V-I>1.5$~mag) the $I$-band
photometry had to be additionally corrected according to formula
presented in Szyma\'nski \etal (2011). For eclipsing variables without
any measurement in $V$ (\ie mostly faint objects with $I>19$~mag) the
$V-I$ value in the color term of the transformation was derived from
a linear extrapolation of the $V-I$ \vs $I$ relation for stars
with $17<I<19$~mag in the same subfield.


\Section{Completeness of the Search}

While preparing the final list of eclipsing binaries 
we searched for the same variables detected in overlapping regions.
Within the largest overlapping region between fields CAR109
and CAR115 we found 206 common stars out of a total of 308 eclipsing
variables detected in this area. This gives an upper limit for the
completeness of the sample of $2x/(x+1)\approx80$\%, where $x=206/308$.
However, a more realistic estimate of the completeness of our search
is based on a comparison with an independent variability survey.

For four contiguous nights in April 2005 a 0.052~deg$^2$ area located
within the OGLE-III disk field CAR105 was monitored to follow-up
selected OGLE transiting candidates (Udalski \etal 2002abc, 2003).
A dataset consisting of 660 $V$-band images was taken with VIMOS of the
European Southern Observatory 8.2-m Very Large Telescope (VLT). Of nine
OGLE transits in the VIMOS area, two objects were later confirmed to be
caused by planets: OGLE-TR-111b (Pont \etal 2004) and OGLE-TR-113b (Bouchy
\etal 2004, Konacki \etal 2004). A search for variables brought 348 objects
among 50 897 stars in the brightness range between $V$=15.4~mag and $V$=24.5~mag
(Pietrukowicz \etal 2009). Despite different bands used in the OGLE and
VIMOS surveys, the depth is very similar or even slightly shallower
in the case of OGLE. Both surveys are very complementary. While VIMOS
observations have an excellent sampling of $\approx165$ exposures per night,
OGLE collected about 1800 points in 167 nights over four seasons.
We cross-matched the list of VIMOS variables with the list of OGLE
eclipsing stars. Eighty-nine objects were common, 23 OGLE variables
were not found in the VIMOS list, while 38 VIMOS eclipsing binaries
were not detected during the OGLE search. Based on these numbers we
conclude that the completeness of the catalog of eclipsing binaries
is $(89+23)/(89+23+38)\approx75$\%. Thirty-four of the missing VIMOS
binaries were added to our list.

The final list of eclipsing variables was additionally extended by 139 OGLE
transits detected in the beginning of the third phase of the project
and reported in Udalski \etal (2002abc, 2003, 2008) and Pont \etal (2008).
Out of a total number of 155 transit candidates reported in the Galactic
disk five transits were spectroscopically confirmed as being due
to planets: OGLE-TR-111b (Pont \etal 2004), OGLE-TR-113b (Bouchy \etal 2004,
Konacki \etal 2004), OGLE-TR-132b (Bouchy \etal 2004), OGLE-TR-182b
(Pont et al. 2008), and OGLE-TR-211b (Udalski \etal 2008). The remaining
stars but five objects turned out to be either binary systems, multiple
systems or systems with unverified planetary companions.
Six of the systems were independently found in our variability search.
With the longer database we have improved the orbital periods for most
of the 145 OGLE transit objects.


\Section{The Catalog of Eclipsing Binaries\\ in the OGLE-III Disk Fields}

The OGLE-III catalog of eclipsing binaries in the Galactic disk,
containing tables with basic parameters, time-series $I$- and $V$-band
photometry, and finding charts, is available to the astronomical
community from the OGLE Internet Archive:
\begin{center}
{\it http://ogle.astrouw.edu.pl\\
ftp://ftp.astrouw.edu.pl/ogle/ogle3/OIII-CVS/gd/ecl/\\}
\end{center}
Eclipsing variables in the Galactic disk are arranged according to
increasing right ascension and named as OGLE-GD-ECL-NNNNN, where
NNNNN is a five-digit consecutive number. Besides coordinates of the
binaries, their orbital periods, and period uncertainties, we also provide
information on the maximum brightness in $I$, the $I$-band amplitude,
the $V-I$ color, the moment of the primary minimum, classification
to contact or non-contact type, and a cross-match identification name
with the VIMOS survey (Pietrukowicz \etal 2009, 2012) and the list
of OGLE-II eclipsing binaries in Carina (Huemmerich and Bernhard 2012),
if exists.


\Section{X-ray counterparts}

Close binaries are often chromospherically active systems, hence they can
emit intense coronal X-ray radiation. In the optical range, the characteristic
features of such systems are sporadic flares observed over timescales of
minutes (\eg Osten \etal 2012) and relatively large variations in the
out-of-eclipse brightness ($\lesssim0.5$~mag) attributed to spot activity
(\eg Udalski \etal 2012, Ta\c{s} \etal 2013, Rozyczka \etal 2013).

We conducted a search for eclipsing binary counterparts to 220 X-ray sources
located within the OGLE-III Galactic disk area with the help of
the X{\footnotesize AMIN} system of the High Energy Astrophysics Science
Archive Research Center\footnote{http://heasarc.gsfc.nasa.gov/}. The sources
were detected by diverse satellite X-ray observatories and of different
angular resolution: Chandra ($\sigma_{\rm d}=0\zdot\arcs6$ at 90\% uncertainty
circle), XMM-Newton ($1\zdot\arcs5$ for bright and $2\arcs$--$4\arcs$ for
faint sources), and ROSAT ($\sim6\arcs$). We searched for optically variable
counterparts within a radius of $18\zdot\arcs2$ (corresponding to 70~pix of
the OGLE-III camera) around the X-ray sources. Two Chandra X-ray sources,
X104523.77-603051.7 and X110523.68-610822.2, coincide with relatively
bright eclipsing binaries OGLE-GD-ECL-02102 ($I_{max}$=14.23~mag) and
OGLE-GD-ECL-06924 ($I_{max}$=13.71~mag), respectively. The angular distance
in the sky between the coincided optical and X-ray positions is only $0\zdot\arcs78$
and $0\zdot\arcs23$, respectively. Moreover, large variations observed in the
light curve of OGLE-GD-ECL-06924 (shown in Fig.~12 in Section 7.4) is a strong
evidence for the ideal X-ray-optical cross-match in this case. One XMM-Newton
source, J104704.2-620158, located $5\zdot\arcs1$ from OGLE-GD-ECL-02579
and two ROSAT sources, J1105.0-6116 and J1155.4-6211, located $3\zdot\arcs6$
from OGLE-GD-ECL-06854 and $5\zdot\arcs0$ from OGLE-GD-ECL-08457, respectively,
are counterpart candidates for these eclipsing systems. For another two ROSAT
X-ray sources (J1105.7-6054 and J115219.1-615605) we found nearby eclipsing
variables (OGLE-GD-ECL-06974 and OGLE-GD-ECL-08046, respectively). However,
the distances between the optical and X-ray positions of $2.05\sigma_{\rm d}$
and $2.78\sigma_{\rm d}$, respectively, make the cross-identification less
certain.


\Section{Brightness, Amplitude, and Period Distributions}

In Fig.~3, we present the observed out-of-eclipse $I$-band brightness
distribution for all binary stars found in the OGLE-III disk area.
For stars with $I>18$~mag the completeness of the sample drops significantly.
Fig.~4 shows the $I$-band amplitude distributions for all detected eclipsing
variables and variables brighter than $I=18$~mag. From this plot we can infer
that the sample of stars with $I<18$~mag lacks systems with amplitudes $<0.2$~mag.

\begin{figure}[htb]
\centerline{\includegraphics[angle=0,width=130mm]{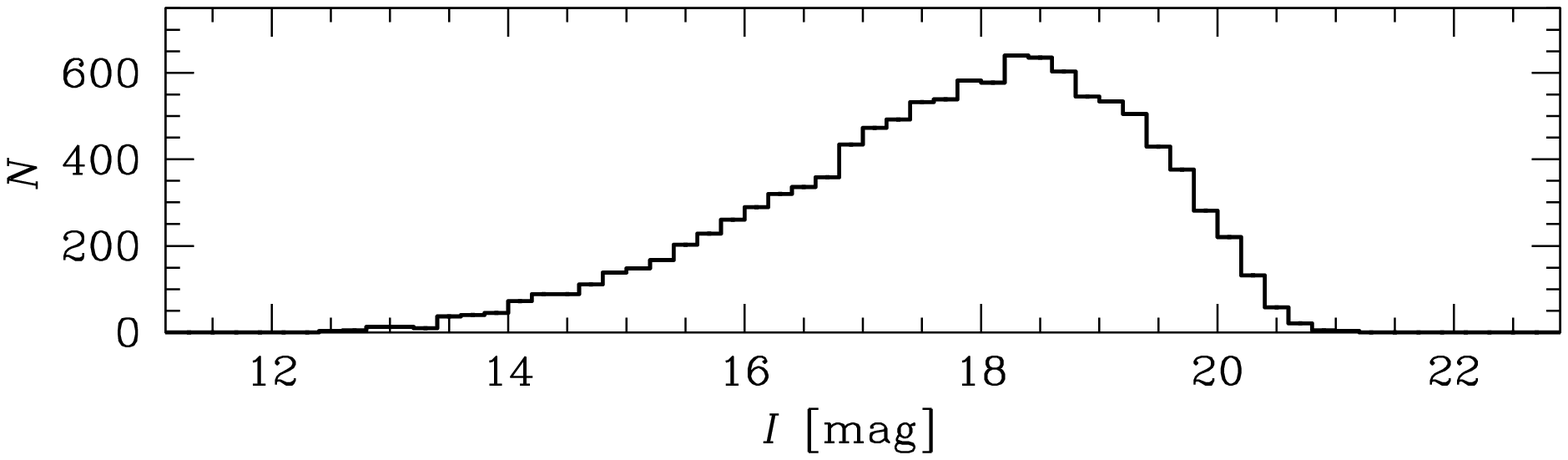}}
\FigCap{Brightness distribution in the $I$ band for 11 589 eclipsing variables
detected in the OGLE-III Galactic disk fields. The bin size is 0.2~mag.}
\end{figure}

\begin{figure}[htb]
\centerline{\includegraphics[angle=0,width=130mm]{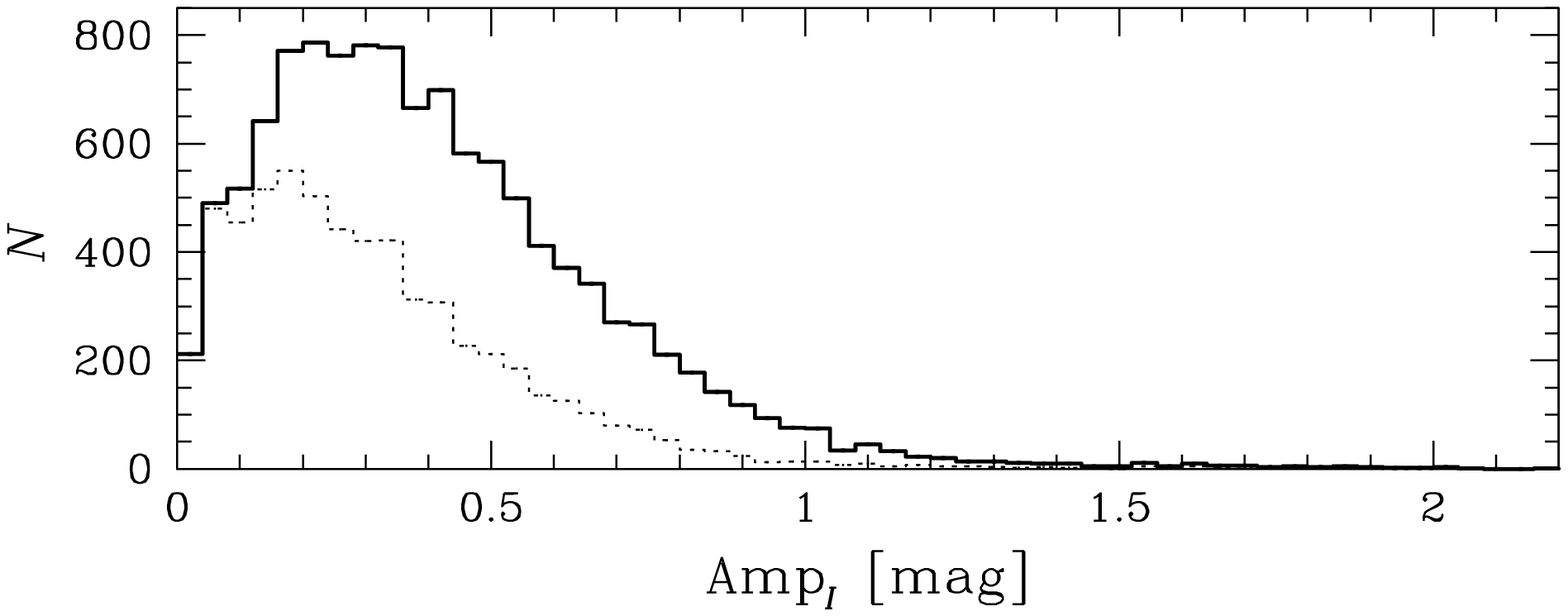}}
\FigCap{Amplitude distributions for all 11 589 OGLE-III disk eclipsing variables
(thick solid line) and 6023 variables with $I<18$~mag (dotted line). The bin
size is 0.04~mag.}
\end{figure}

In the next figure, Fig.~5, we plot the distribution of the orbital
period for the OGLE-III disk binaries. The shortest orbital period
of $0.07753698(1)$~d was found in a hot subdwarf binary star
OGLE-GD-ECL-10384, while the longest orbital period of 103.502(1)~d
was measured for system OGLE-GD-ECL-03367. Based on the shape of the light
curves we divided our whole sample into likely contact and non-contact
systems. Contact systems constitute about 64\% of all detected binaries. Period
distributions for the two separate groups are also shown in Fig.~5. It is
noticable that the majority of binaries with $P<0.5$~d are in contact. Among
binaries with orbital periods around 0.7~d about half are contact systems.
Their number significantly decreases around $P\approx1.5$~d, although some
of them may have longer periods. The shortest orbital period of a binary,
classified as a probable contact system, equals to 0.1571319(2)~d in the case
of OGLE-GD-ECL-00665, while the longest period of 19.6464(2)~d was found
in a contact system candidate OGLE-GD-ECL-03668. We have to stress that
our classification is tentative. Contact system candidates with
unusually short and unusually long periods are not confirmed and require
more studies.

\begin{figure}[htb]
\centerline{\includegraphics[angle=0,width=130mm]{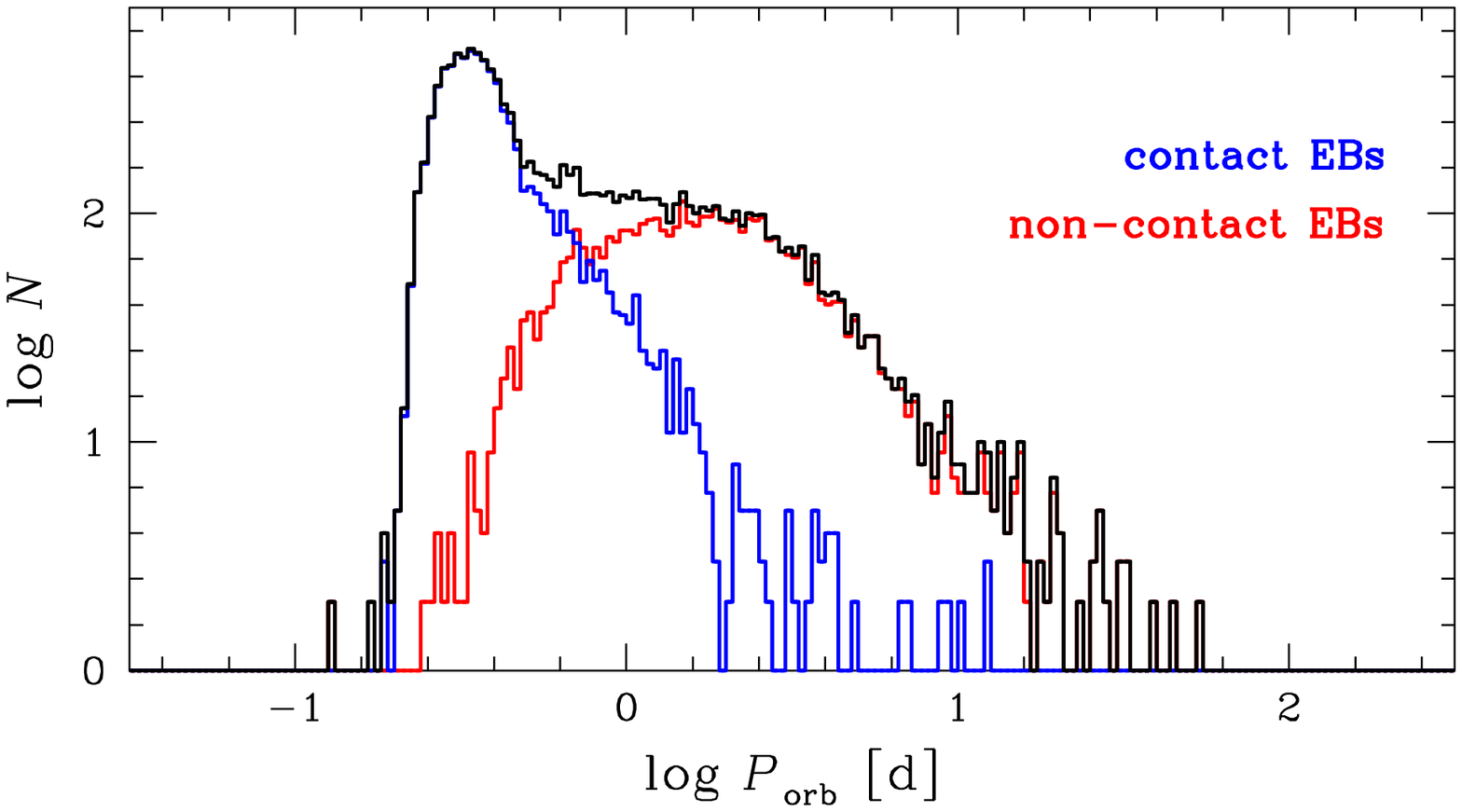}}
\FigCap{Orbital period distribution for 11 562 OGLE-III disk eclipsing
binaries (EBs) with a division into contact and non-contact systems.
Note that the histogram shows the logarithm of the counts, thus bins
with a single star appear blank.}
\end{figure}

The orbital period distribution for 6023 OGLE-III disk binaries
with brightness $I<18$~mag is shown in Fig.~6. In the same figure,
we also draw the orbital period distributions for eclipsing binaries
discovered in the course of two recent wide-field surveys: the All Sky
Automated Survey (ASAS, Paczy\'nski et al. 2006) and the Kepler space
telescope (Slawson et al. 2011). The ASAS survey monitored stars
with brightness $V\lesssim14$~mag and declination $<+28\arcd$ for 5--8
years (Pojma\'nski 1997, 2002, 2003). The presented period distribution
is based on 5462 ASAS variables unambiguously classified as eclipsing
binaries with $V<12$~mag. The Kepler telescope monitored a fixed area
of 115~deg$^2$ toward mainly constellations of Cygnus and Lyra
between Galactic latitudes $+6\arcd$ and $+23\arcd$ in searches for
transiting extrasolar planets (\eg Koch \etal 2010). The plotted
orbital period distribution is based on 1032 binaries with Kepler
magnitudes $<14$~mag which were identified in a dataset of continuous
observations lasting 125~days.

Although the three surveys, OGLE, ASAS, and Kepler, observed different
Galactic regions and different stellar populations, we can draw common
conclusions from the comparison presented in Fig.~6. First, the
distribution of the orbital period peaks around 0.40~d. Obviously,
this maximum refers to contact binaries. Second, for orbital periods
between 0.5~d and 2.5~d the distribution is either flat or decreases very
weakly with $P$. Linear fit to the rich OGLE data in this regime does
not give us a clear answer: log$N\propto(-0.025\pm0.036){\rm log}P$.
For periods longer than 2.5~days the Kepler observations indicate
a higher occurance rate of binaries than more numerous and longer-duration
ground-based observations from OGLE and ASAS. This period regime requires
more data.

\begin{figure}[htb]
\centerline{\includegraphics[angle=0,width=130mm]{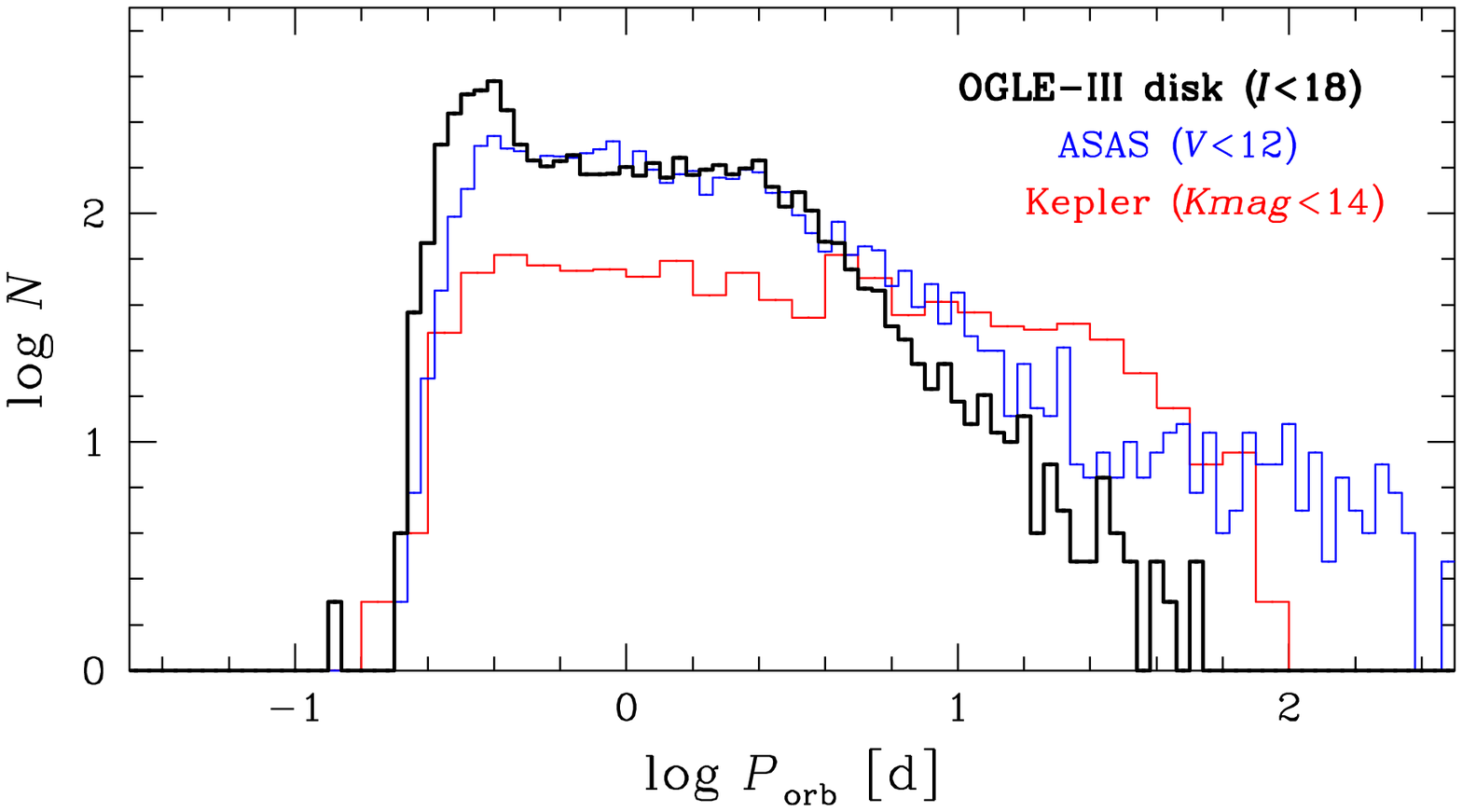}}
\FigCap{Orbital period distribution for bright eclipsing binaries
from the OGLE-III disk fields in comparison with results from
the All-Sky Automatic Survey and the Kepler orbiting telescope.}
\end{figure}


\Section{Interesting Objects}

\Subsection{Candidates for Double Binaries}

Eleven single detections showing light variations due to the presence
of two binary systems were found among thousands of eclipsing objects
discovered in the OGLE-III disk fields. We subtracted the stronger
component with the period $P_a$ from each of the eleven light
curves using Fourier series and searched for the second period $P_b$.
By subtracting the light variations with $P_b$ from the original
light curve we improved the value of $P_a$. Resolved light curves
are plotted in Figs.~7--9.

We made an attempt to verify photometrically if the double binaries either are
likely blended objects observed almost exactly in the same line of sight or
may form physically bound systems. In order to check that we measured
centroid positions of these objects on images with the help of the
DO{\footnotesize PHOT} software (Schechter, Saha and Mateo 1993). Then,
separately in $x$ and $y$ directions, we phased the offsets from the mean
position with periods $P_a$ and $P_b$. The results are also included
in Figs.~7--9, where we binned the measured offsets in $0.01P$ bins. One
can notice that the position changes in OGLE-GD-ECL-00259b and OGLE-GD-ECL-04406a
are correlated with the component's light curves. This may indicate
that nine of the observed double binaries (OGLE-GD-ECL-03436, 05310, 05390,
05656, 07057, 07157, 07443, 10263, 11021) are physically bound quadruple
systems, while in the case of OGLE-GD-ECL-00259 and OGLE-GD-ECL-04406
the component binaries seem to be located almost along the same line of sight
by pure chance.

The probability of such a chance position of two binaries is very small.
The number of 11 589 binary stars detected over the whole OGLE-III
disk area of 7.12~deg$^2$ gives on average 0.000126 objects
per 1.0 arcsec$^2$. At a separation of the mean seeing of the
$I$-band images of $1\zdot\arcs2$ (Szyma\'nski \etal 2011)
the expected number of eclipsing objects would be 0.00018.

\begin{figure}[htb]
\centerline{\includegraphics[angle=0,width=130mm]{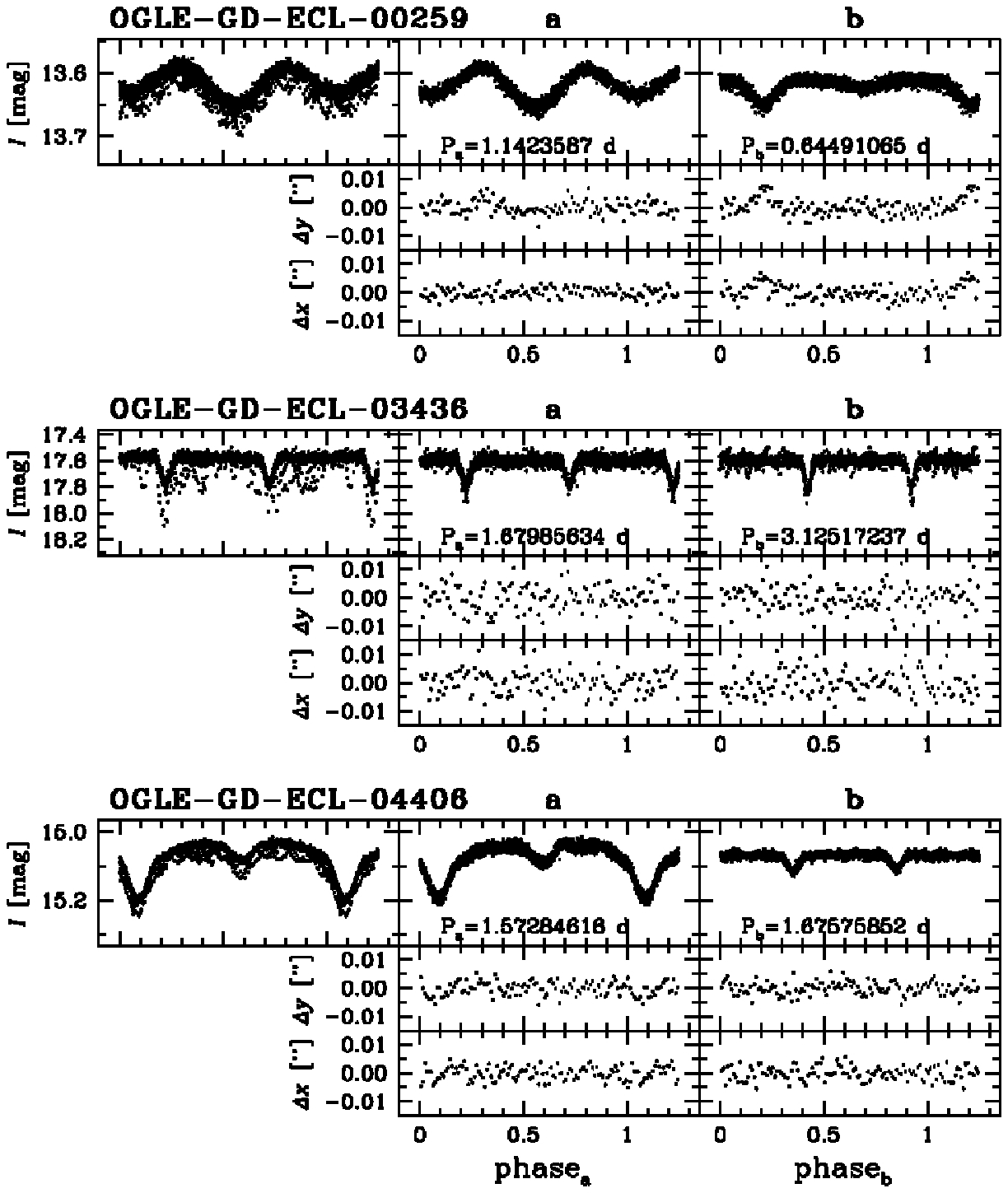}}
\FigCap{Phased light curves of double binaries (left panels).
Each component light curve ('a' and 'b', middle and right panels,
respectively) is phased with the given period ($P_a$ and $P_b$).
The panels below the light curves show offsets in the position
of centriods in the $x$ and $y$ directions of the CCD detector phased
with the same periods. Clear correlation is seen in OGLE-GD-ECL-00259b
between the phased light curve and offsets from the mean position
in both directions and indicates high probability of chance alignment
of the ellipsoidal binary OGLE-GD-ECL-00259a and the eclipsing
binary OGLE-GD-ECL-00259b in the sky. A similiar correlation occurs
for component 'a' of OGLE-GD-ECL-04406. In the case of OGLE-GD-ECL-03436
no such clear correlation is seen, rising the probability of a physical
relation between the two binaries.}
\end{figure}

\begin{figure}[htb]
\centerline{\includegraphics[angle=0,width=130mm]{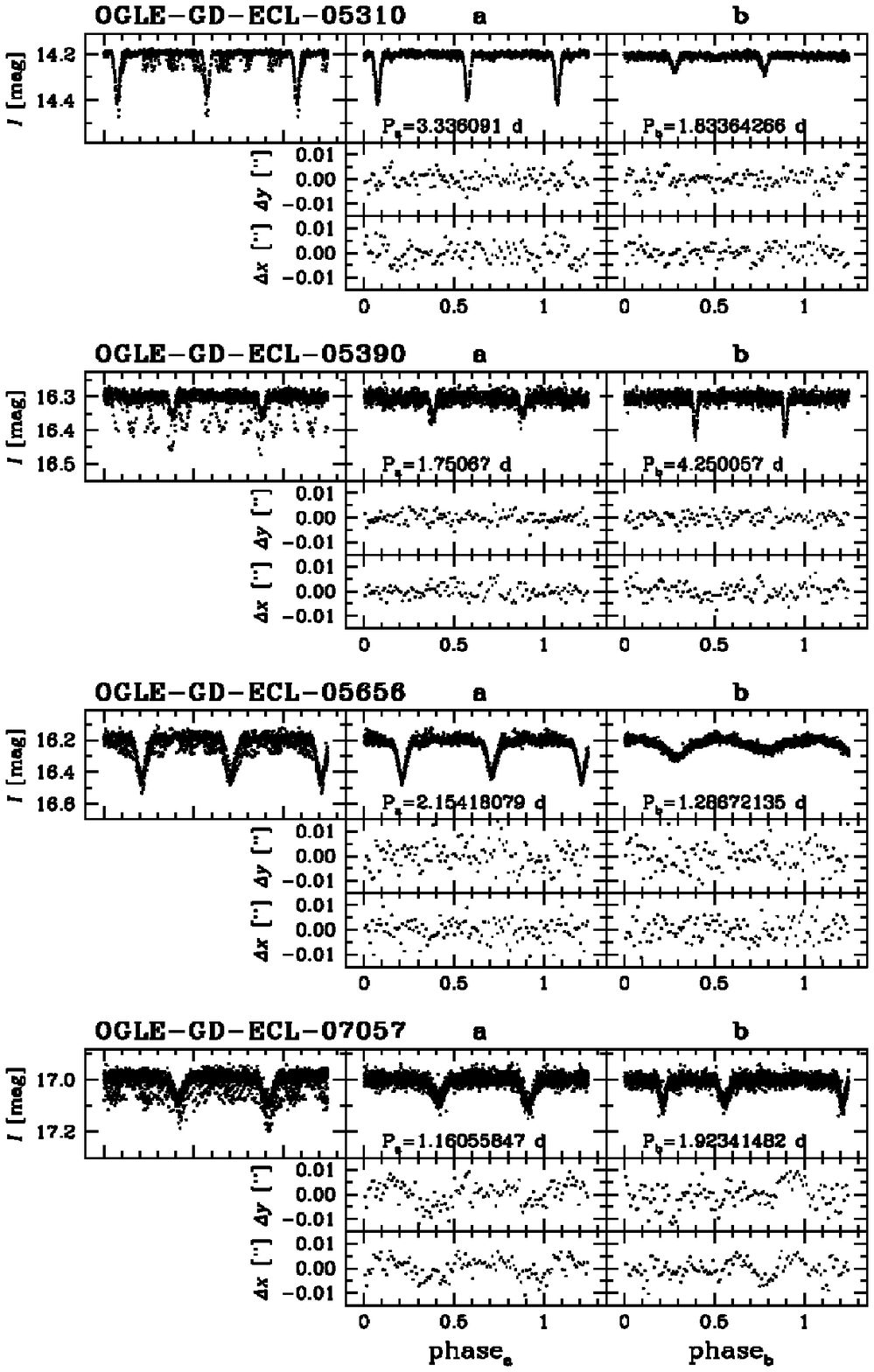}}
\FigCap{Phased light curves of double binaries: OGLE-GD-ECL-05310,
OGLE-GD-ECL-05390, OGLE-GD-ECL-05656, OGLE-GD-ECL-07057.
All four doubly eclipsing binaries are likely quadruple systems.}
\end{figure}

\begin{figure}[htb]
\centerline{\includegraphics[angle=0,width=130mm]{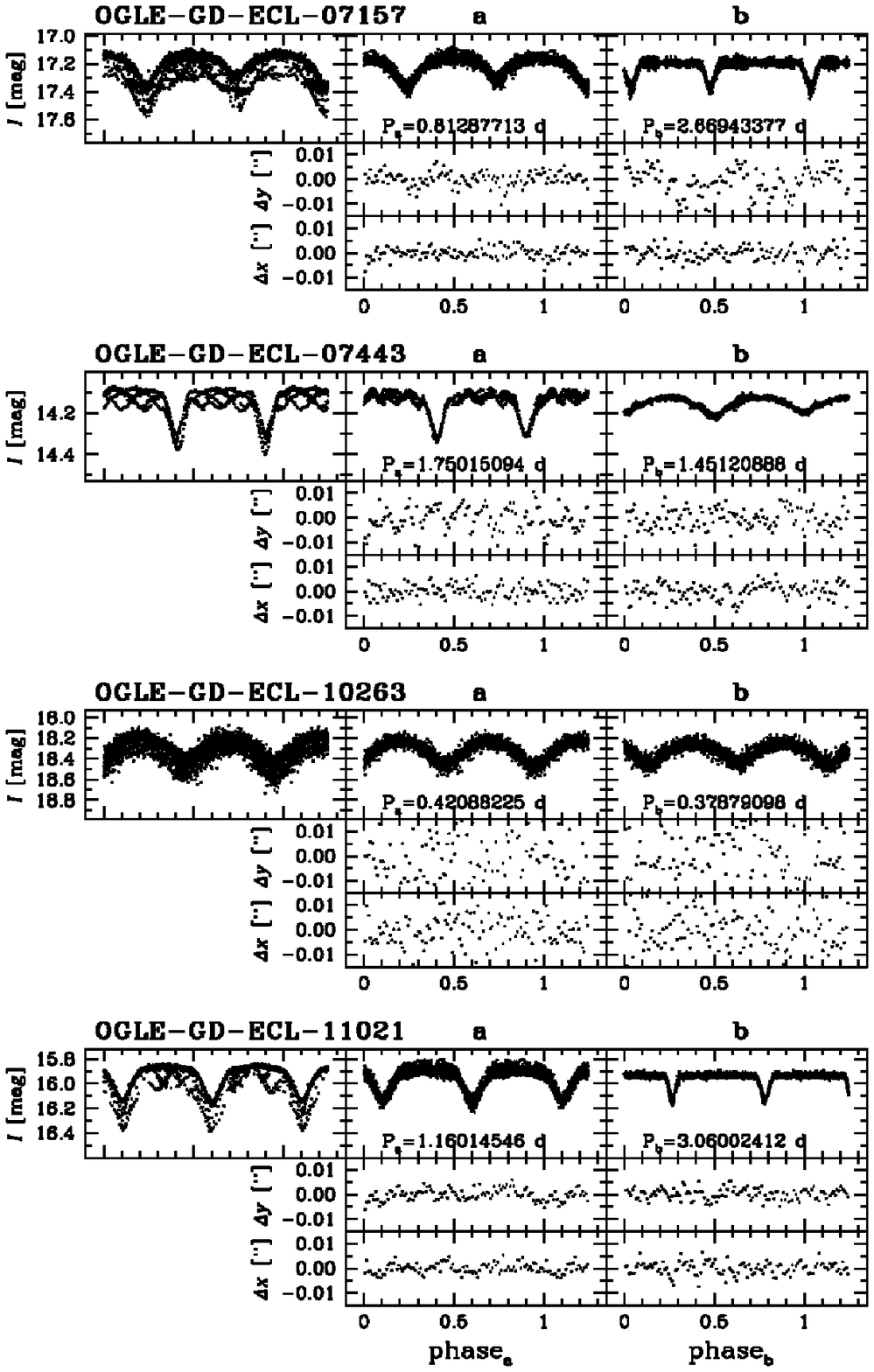}}
\FigCap{Phased light curves of double binaries: OGLE-GD-ECL-07157,
OGLE-GD-ECL-07443, OGLE-GD-ECL-10263, OGLE-GD-ECL-11021.
All four doubly eclipsing binaries are likely quadruple systems.
In the light curve of OGLE-GD-ECL-07443a, one can notice additional
variations. After subtraction of the main signal with $P_a=1.75015094$~d
we obtained a modulation with a period of 0.290005~d, probably due
to spots on the surface of one of the stars.
}
\end{figure}

\Subsection{Subdwarf-B Type Binaries}

In our search for binaries in the disk data
we identified ten binaries with hot components manifestating
their presence in the reflection effect and huge difference
in depth of the two minima (see light curves in Fig.~10).
The binaries have periods between 0.0775~d and 0.5066~d
and five of them are the shortest period systems detected
in the whole sample. These stars are very likely subdwarf
B type (sdB) binaries.

\begin{figure}[htb]
\centerline{\includegraphics[angle=0,width=130mm]{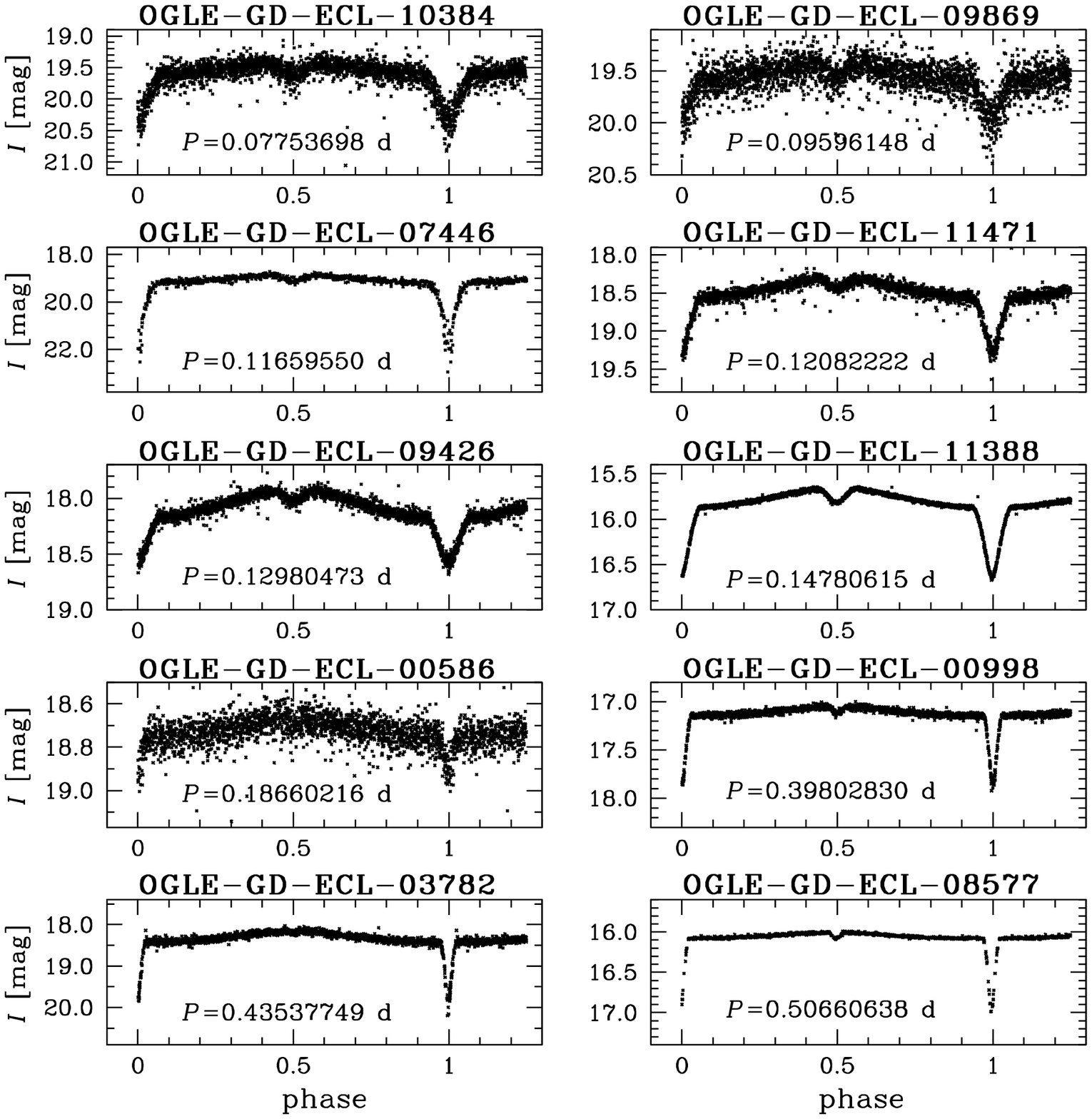}}
\FigCap{Light curves of ten newly discovered sdB type binaries
arranged with the increasing orbital period.}
\end{figure}

\Subsection{Eclipsing RS CVn Type Stars}

RS CVn variable stars are close binaries with
active chromospheres. It is belived that brightness variations
$\lesssim0.5$~mag observed in this type of stars are caused by
huge cool spots on the surface of the components. The high activity
of the system reveals also in sporadic flares and X-ray radiation.
In some systems, signs of accretion are seen (\eg Rozyczka \etal 2013,
Kang \etal 2013). The presence of eclipses in a RS CVn type star
helps to investigate the relation between its activity and the
evolutionary stage of the binary. In Fig.~11, we present the best
examples of RS~CVn type stars identified in the OGLE sample.
As we mentioned in Section 5, one of the presented systems,
OGLE-GD-ECL-06924, has an X-ray counterpart.

\begin{figure}[htb]
\centerline{\includegraphics[angle=0,width=130mm]{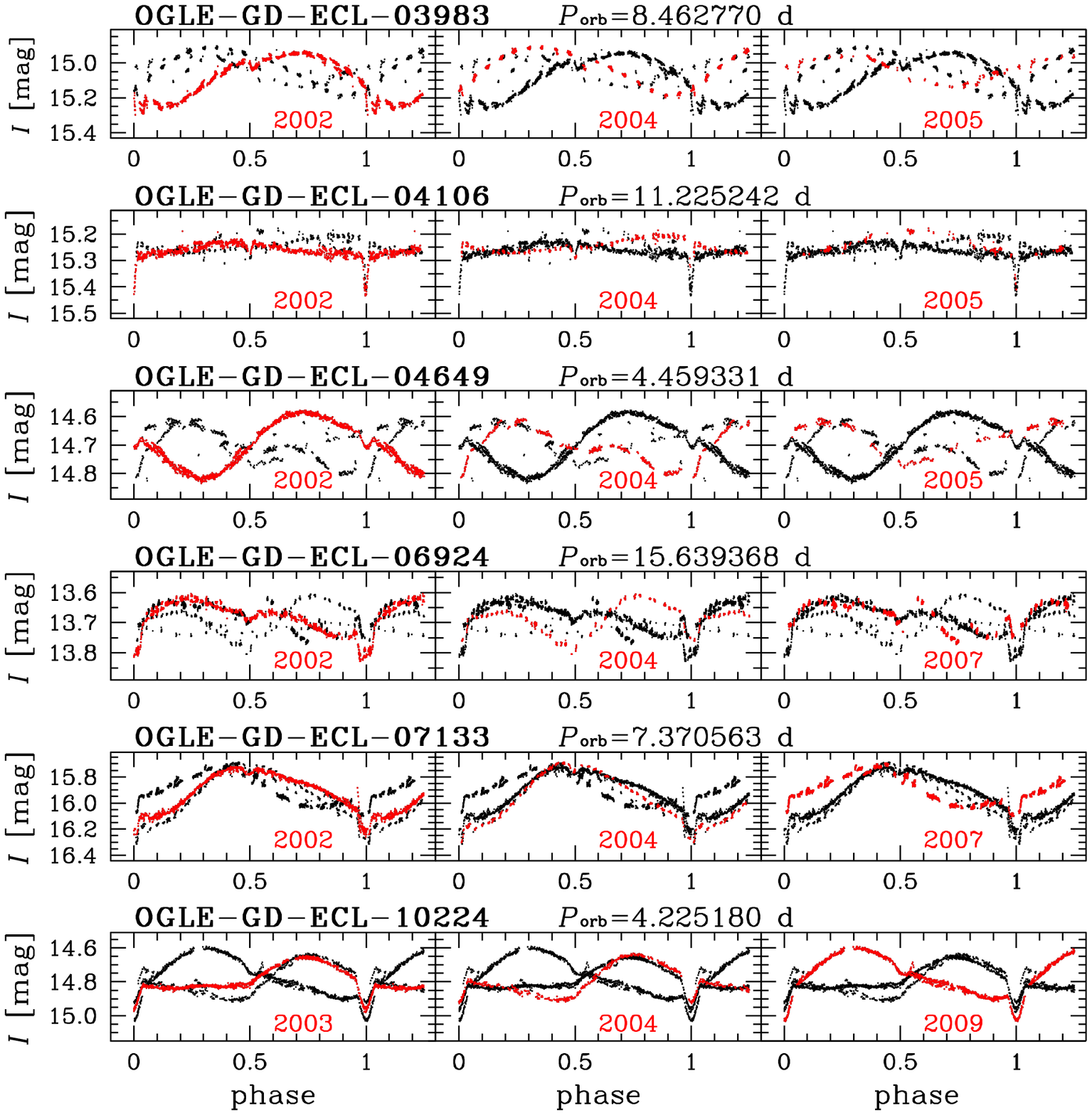}}
\FigCap{Example light curves of six eclipsing RS~CVn type stars
identified in the OGLE-III disk area. The light curves are phased
with the given orbital period. The changing starspot wave is highlighted
in red for three best-covered seasons. Variable OGLE-GD-ECL-04649
shows both a single and a double starspot wave at variuos seasons,
indicating the presence of two dark spot regions on opposite
hemispheres.}
\end{figure}

\Subsection{Eclipsing Objects of Unknown Nature}

In Fig.~12, we present phased light curves of three eclipsing variables
whose nature is unknown. The objects have amplitudes between 0.13~mag and 0.27~mag
in the $I$ band and the observed $V-I$ color in the range from 2.97~mag to 3.38~mag.
The asymmetric eclipse in OGLE-GD-ECL-03166 may indicate the presence of
an accretion disk. Object OGLE-GD-ECL-06674 has a very short period of
0.13004370(7)~d, typical for close compact binaries such as polars and
intermediate polars. Unfortunately, there are no detected X-ray counterparts
to these stars, what could help in their final identification.

\begin{figure}[htb]
\centerline{\includegraphics[angle=0,width=130mm]{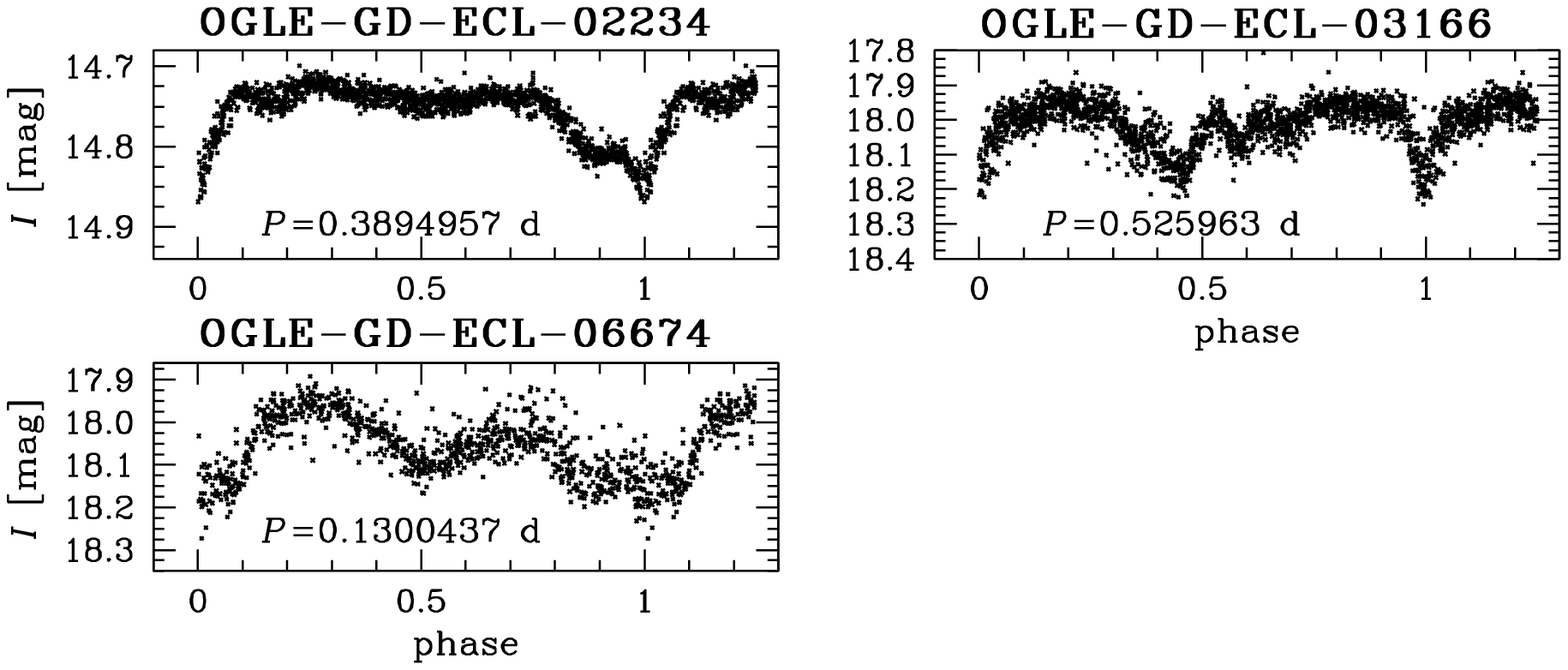}}
\FigCap{Phased light curves of three eclipsing objects of unknown nature.}
\end{figure}


\Section{Summary and conclusions}

In this paper we report the indentification of 11 589 eclipsing objects
detected in the OGLE-III Galactic disk fields. All objects but 402
variable stars are new discoveries. The completeness of our catalog
we estimate to be at a level of 75\%. We tentatively classified the
eclipsing stars into contact and non-contact systems. The first group
constitutes about two-thirds of all identified systems. The orbital
period distribution of the OGLE binaries shows a maximum at $\approx0.40$~d
and an almost flat part between 0.5~d and 2.5~d. After comparison with results
from the ASAS and Kepler data, we find that these two properties
seem to be independent of population. A conclusion from the obtained period
distribution is that binaries tend to shrink their orbits.
This is in agreement with the well known fact of the mass and momentum
loss during the evolution of binaries.

In Fig.~13, we plot an interesting finding based on our data: the ratio
of non-contact to contact systems for latitudes $b<-1\zdot\arcd5$
seems to be constant around 0.5, while it increases up to 1.0 around
$b\sim-1\zdot\arcd0$. This means that close to the Galactic plane there
is a deficiency of systems in contact. There is a possibility that some
of the missing systems could have merged, as it was observed in the case
of the red nova V1309 Sco (Tylenda \etal 2011).

\begin{figure}[htb]
\centerline{\includegraphics[angle=0,width=130mm]{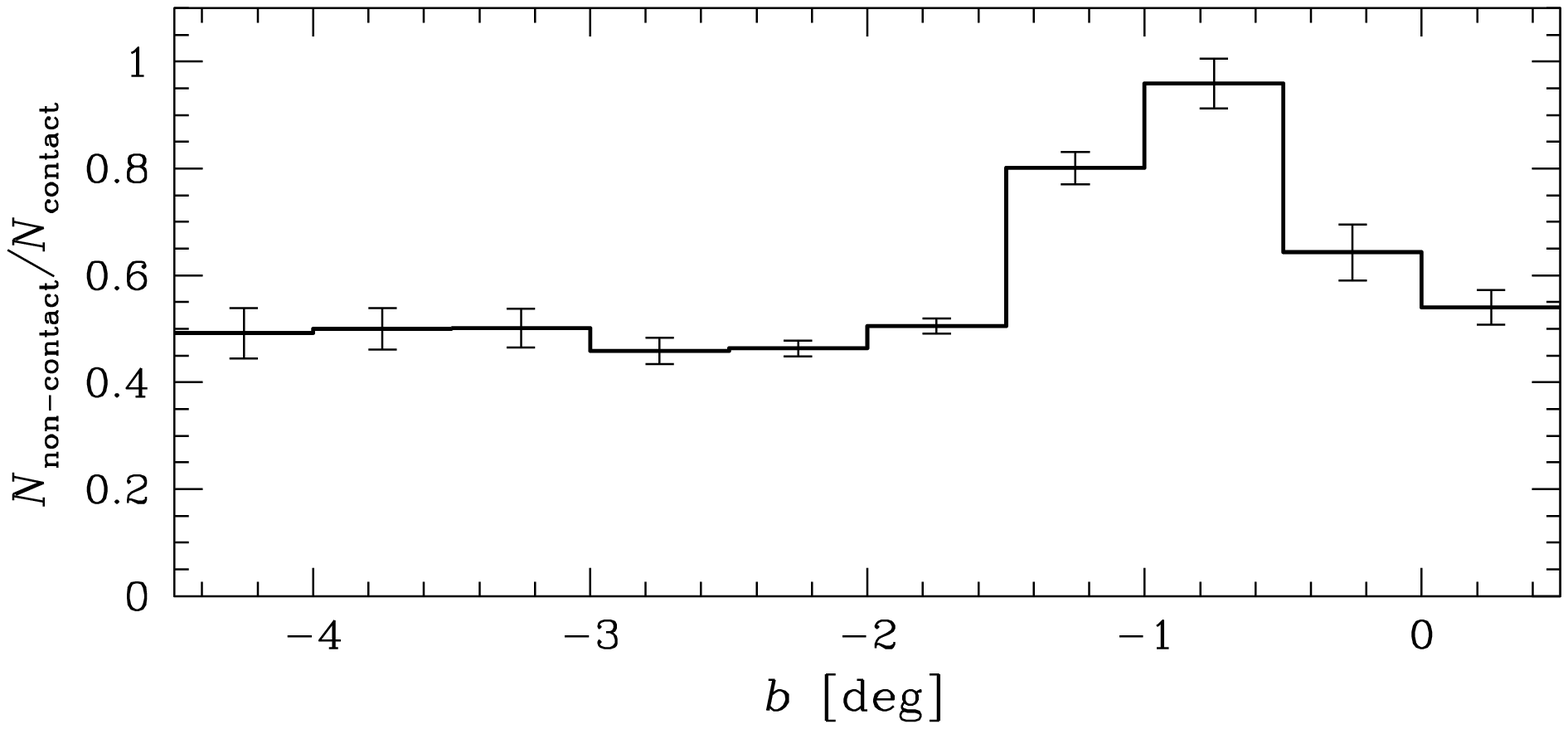}}
\FigCap{Noncontact-to-contact binary ratio as a function of the Galactic
latitude. Note a significant increase of the ratio around latitude -1\arcd.}
\end{figure}

Eleven single unresolved detections showing light variations due to the
presence of two binary systems were found among thousands of eclipsing
objects. Astrometric analysis indicates that nine of the objects are good
candidates for quadruple systems, while the other two are likely
unresolved blends of two unbound binary systems. Our search for
eclipsing objects brought identification of ten subdwarf-B type binaries
and dozens of RS~CVn type variables. We also found three objects
with eclipses of unknown origin. All these objects need multi-band
and spectroscopic follow-up studies.

\Acknow{
This work has been supported by the Polish National Science Centre
grant No. DEC-2011/03/B/ST9/02573 and the Polish Ministry of Sciences
and Higher Education grant No. IP2012 005672 under Iuventus Plus
programme. The OGLE project has received funding from the European
Research Council under the European Community$'$s Seventh Framework
Programme (FP7/2007-2013)/ERC grant agreement No. 246678 to A.U.}



\begin{references}

\refitem{Alard, C., and Lupton, R. H.}{1998}{\ApJ}{503}{325}
\refitem{Bonanos, A. Z., \etal}{2006}{\ApJ}{652}{313}
\refitem{Bouchy, F., Pont, F., Santos, N. C., Melo, C., Mayor, M., Queloz, D., Udry, S.}{2004}{\AA}{421}{L13}
\refitem{Doyle, L. R., \etal}{2011}{Science}{333}{1602}
\refitem{Graczyk, D., Soszy\'nski, I., Poleski, R., Pietrzy\'nski, G., Udalski, A., Szyma\'nski, M. K., Kubiak, M., Wyrzykowski, {\L}., and Ulaczyk, K.}{2011}{\Acta}{61}{103}
\refitem{Huemmerich, S., and Bernhard, K.}{2012}{Peremennye Zvezdy Prilozhenie}{12}{11}
\refitem{He{\l}miniak, K. G., Devor, J., Minniti, D., and Sybilski, P.}{2013}{\MNRAS}{432}{2895}
\refitem{Kang, Y.-W., Yushchenko, A. V., Hong, K., Guinan, E. F., and Gopka, V. F.}{2013}{\AJ}{145}{167}
\refitem{Koch, D. G. \etal}{2010}{\ApJ}{713}{L79}
\refitem{Konacki, M., Torres, G., Sasselov, D. D., Pietrzy\'nski, G., Udalski, A., Jha, S., Ru\'iz, M. T., Gieren, W., Minniti, D.}{2004}{\ApJ}{609}{L37}
\refitem{Morales, J. C., Gallardo, J., Ribas, I., Jordi, C., Baraffe, I., and Chabrier, G.}{2010}{\ApJ}{718}{502}
\refitem{Mr\'oz, P, Pietrukowicz, P., Soszy\'nski, I., Udalski, A., Poleski, R., Szyma\'nski, M. K., Kubiak, M., Pietrzy\'nski, G., Wyrzykowski, {\L}., Ulaczyk, K., Koz{\l}owski, S., and Skowron, J}{2013}{\Acta}{63}{NNN}
\refitem{Nataf, D. M., Gould, A., and Pinsonneault, M. H.}{2012}{\Acta}{62}{33}
\refitem{North, P., Gauderon, R., Barblan, F., and Royer, F.}{2010}{\AA}{520A}{74}
\refitem{Osten, R. A., Kowalski, A., Sahu, K., and Hawley, S. L.}{2012}{\ApJ}{754}{4}
\refitem{Paczy\'nski, B., Szczygie{\l}, D. M., Pilecki, B., and Pojma\'nski, G.}{2006}{\MNRAS}{368}{1311}
\refitem{Pietrukowicz, P., Minniti, D., Fern\'andez, J. M., Pietrzy\'nski, G., Ru\'iz, M. T., Gieren, W., D\'iaz, R. F., Zoccali, M., and Hempel, M.}{2009}{\AA}{503}{651}
\refitem{Pietrukowicz, P., Minniti, D., Alonso-Garc\'ia, J., and Hempel, M.}{2012}{\AA}{537}{A116}
\refitem{Pietrzy\'nski, G. \etal}{2010}{Nature}{468}{542}
\refitem{Pietrzy\'nski, G. \etal}{2012}{Nature}{484}{75}
\refitem{Pietrzy\'nski, G. \etal}{2013}{Nature}{495}{76}
\refitem{Pojma\'nski, G.}{1997}{\Acta}{47}{467}
\refitem{Pojma\'nski, G.}{2002}{\Acta}{52}{397}
\refitem{Pojma\'nski, G.}{2003}{\Acta}{53}{341}
\refitem{Pont, F., Bouchy, F., Queloz, D., Santos, N. C., Melo, C., Mayor, M., Udry, S.}{2004}{\AA}{426}{L15}
\refitem{Pont, F. \etal}{2008}{\AA}{487}{749}
\refitem{Ribas, I., Jordi, C., and Gim\'enez, \'A.}{2000}{\MNRAS}{318}{L55}
\refitem{Rozyczka, M., Pietrukowicz, P., Kaluzny, J., Pych, W., Angeloni, R., D\`ek\'any, I.}{2013}{\MNRAS}{429}{1840}
\refitem{Schechter, P. L., Saha, K., and Mateo, M.}{1993}{\PASP}{105}{1342}
\refitem{Schwarzenberg-Czerny, A.}{1996}{\ApJ}{460}{L107}
\refitem{Slawson, R. W., Pr\v sa, A., Welsh, W. F., Orosz, J. A., Rucker, M., Batalha, N. \etal}{2011}{\AJ}{142}{160}
\refitem{Soszy\'nski, I., Poleski, R., Udalski, A., Szyma\'nski, M. K., Kubiak, M., Pietrzy\'nski, G., Wyrzykowski, {\L}., Szewczyk, O., and Ulaczyk, K.}{2008}{\Acta}{58}{163}
\refitem{Soszy\'nski, I., Dziembowski, W. A., Udalski, A., Poleski, R., Szyma\'nski, M. K., Kubiak, M., Pietrzy\'nski, G., Wyrzykowski, {\L}., Ulaczyk, K., Koz{\l}owski, S., and Pietrukowicz, P.}{2011}{\Acta}{61}{1}
\refitem{Soszy\'nski, I., Udalski, A., Szyma\'nski, M. K., Kubiak, M., Pietrzy\'nski, G., Wyrzykowski, {\L}, Ulaczyk, K., Poleski, R., Koz{\l}owski, S., Pietrukowicz, P., and Skowron, J.}{2013}{\Acta}{63}{21}
\refitem{S\"oderhjelm, S. and Dischler, J.}{2005}{\AA}{442}{1003}
\refitem{Szyma\'nski, M. K., Udalski, A., Soszy\'nski, I., Kubiak, M., Pietrzy\'nski, G., Poleski, R., Wyrzykowski, {\L}., and Ulaczyk, K.}{2010}{\Acta}{60}{295}
\refitem{Szyma\'nski, M. K., Udalski, A., Soszy\'nski, I., Kubiak, M., Pietrzy\'nski, G., Poleski, R., Wyrzykowski, {\L}., and Ulaczyk, K.}{2011}{\Acta}{61}{83}
\refitem{Ta\c{s}, G., and Evren, S.}{2013}{Astr. Nacht.}{334}{251}
\refitem{Tylenda, R., Hajduk, M., Kami\'nski, T., Udalski, A., Soszy\'ski, I., Szyma\'nski, M. K., Kubiak, M., Pietrzy\'nski, G., Poleski, R., Wyrzykowski, {\L}., and Ulaczyk, K.}{2011}{\AA}{528A}{114}
\refitem{Udalski, A., Szyma\'nski, M., Kaluzny, J., Kubiak, M., Krzemi\'nski, W., Mateo, M., Preston, G. W., Paczy\'nski, B.}{1993}{\Acta}{43}{289}
\refitem{Udalski, A., Paczy\'nski, B., \.Zebru\'n, K., Szyma\'nski, M., Kubiak, M., Soszy\'nski, I., Szewczyk, O., Wyrzykowski, {\L}., and Pietrzy\'nski, G.}{2002a}{\Acta}{52}{1}
\refitem{Udalski, A., \.Zebru\'n, K., Szyma\'nski, M., Kubiak, M., Soszy\'nski, I., Szewczyk, O., Wyrzykowski, {\L}., Pietrzy\'nski, G.}{2002b}{\Acta}{52}{115}
\refitem{Udalski, A., Szewczyk, O., \.Zebru\'n, K., Pietrzy\'nski, G., Szyma\'nski, M., Kubiak, M., Soszynski, I., and Wyrzykowski, {\L}.}{2002c}{\Acta}{52}{317}
\refitem{Udalski, A., Pietrzy\'nski, G., Szyma\'nski, M., Kubiak, M., \.Zebru\'n, K., Soszy\'nski, I., Szewczyk, O., and Wyrzykowski, {\L}.}{2003}{\Acta}{53}{133}
\refitem{Udalski, A.}{2003}{\Acta}{53}{291}
\refitem{Udalski, A. \etal}{2008}{\AA}{482}{299}
\refitem{Udalski, A., Szyma\'nski, M. K., Soszy\'nski, I, and Poleski, R.}{2008}{\Acta}{58}{69}
\refitem{Udalski, A., Kowalczyk, K., Soszy\'nski, I., Poleski, R., Szyma\'nski, M. K., Kubiak, M., Pietrzy\'nski, G., Koz{\l}owski, S., Pietrukowicz, P., Ulaczyk, K., Skowron, J., Wyrzykowski, {\L}.}{2012}{\Acta}{62}{133}
\refitem{Vilardell, F., Ribas, I., Jordi, C., Fitzpatrick, E. L., and Guinan, E. F.}{2010}{\AA}{509A}{70}
\refitem{Welsh, W. F., \etal}{2012}{Nature}{481}{475}
\refitem{Wo\'zniak, P. R.}{2000}{\Acta}{50}{421}

\end{references}
\end{document}